\documentclass[conference]{IEEEtran}

\IEEEoverridecommandlockouts
\usepackage{cite}
\usepackage{amsmath,amssymb,amsfonts}
\usepackage{graphicx}
\usepackage{textcomp}
\usepackage{xcolor}
\usepackage{blindtext}
\usepackage[ruled,linesnumbered,lined,boxed,commentsnumbered]{algorithm2e}
\usepackage{caption}
\usepackage{subcaption} 
\usepackage{tabularx}
\usepackage{array}
\usepackage{listings}
\usepackage{booktabs}
\usepackage{makecell}
\usepackage{colortbl}

\usepackage[switch]{lineno}

\newcommand{\NMgemm}{\mathbin{\text{\footnotesize$\bigcirc$\kern-0.75em$\ast$}}}

\def\BibTeX{{\rm B\kern-.05em{\sc i\kern-.025em b}\kern-.08em
    T\kern-.1667em\lower.7ex\hbox{E}\kern-.125emX}}
\begin{document}

\title{
NM-SpMM: Accelerating Matrix Multiplication Using \textit{N:M} Sparsity with GPGPU \\
\author{
    \IEEEauthorblockN{Cong Ma$^{1,2}$, Du Wu$^{3,4}$, Zhelang Deng$^{1,9}$, Jiang Chen$^{1,5}$, Xiaowen Huang$^{6}$, Jintao Meng$^{1*}$, Wenxi Zhu$^{5*}$,\\
   Bingqiang Wang$^{7}$, Amelie Chi Zhou$^{8}$, Peng Chen$^{3}$, Minwen Deng$^{5*}$, Yanjie Wei$^{1}$, Shengzhong Feng$^{9}$, Yi Pan$^{1}$}
    \IEEEauthorblockA{
        \textit{$^1$Shenzhen Institutes of Advanced Technology, Chinese Academy of Sciences, China} \\
        \textit{$^2$University of Chinese Academy of Sciences, China} \\
        \textit{$^3$RIKEN Center for Computational Science, Japan} \\ 
        \textit{$^4$Institute of Science Tokyo, Japan} \\
        \textit{$^5$Tencent AI Lab, China} \\
        \textit{$^6$Shenzhen University, China} \\
        \textit{$^7$Peng Cheng Laboratory, China} \\
        \textit{$^8$Hong Kong Baptist University, China} \\
        \textit{$^9$Guangdong Institute of Intelligence Science and Technolog, China}
    }
}
}

\maketitle

\begin{abstract}

Deep learning demonstrates effectiveness across a wide range of tasks.
However, the dense and over-parameterized nature of these models results in significant resource consumption during deployment. 
In response to this issue, weight pruning, particularly through \textit{N:M} sparsity matrix multiplication, offers an efficient solution by transforming dense operations into semi-sparse ones.
\textit{N:M} sparsity provides an option for balancing performance and model accuracy, but introduces more complex programming and optimization challenges.
To address these issues, we design a systematic top-down performance analysis model for \textit{N:M} sparsity.
Meanwhile, NM-SpMM is proposed as an efficient general \textit{N:M} sparsity implementation. 
Based on our performance analysis, NM-SpMM employs a hierarchical blocking mechanism as a general optimization to enhance data locality, while memory access optimization and pipeline design are introduced as sparsity-aware optimization, allowing it to achieve close-to-theoretical peak performance across different sparsity levels.
Experimental results show that NM-SpMM is 2.1x faster than nmSPARSE (the state-of-the-art for general \textit{N:M} sparsity) and 1.4x to 6.3x faster than cuBLAS’s dense GEMM operations, closely approaching the theoretical maximum speedup resulting from the reduction in computation due to sparsity.
NM-SpMM is open source and publicly available at https://github.com/M-H482/NM-SpMM.

\end{abstract}


\section{Introduction}

Deep learning, especially large language models (LLMs), demonstrates  effectiveness across a wide range of tasks, including computer vision, natural language processing, knowledge representation, recommendation systems, drug discovery, and more~\cite{vaswani2017attention,devlin-etal-2019-bert,radford2019language,brown_llm_few_nips2020,touvron2023llama}.
However, artificial deep learning models are traditionally dense and over-parameterized~\cite{Bengio_sparse_2021,denil2013predicting,Torsten_sparsity_in_dl}, leading to significant consumption of computing and memory resources during actual real-world deployment.
To address this issue, weight pruning serves as an effective strategy for reducing the size of deep learning models by eliminating less relevant weight elements while maintaining model accuracy~\cite{solla1990optimal,han2015learning,gale_sparse_2020,Hassibi_OBS_1993}.
\textit{N:M} sparsity matrix multiplication is the most performant solution in the weight pruning field, where dense matrix multiplications in the model are converted into semi-sparse matrix multiplications~\cite{Torsten_sparsity_in_dl,child2019generating,domingos1999role}.

The optimization of \textit{N:M} sparsity matrix multiplication is motivated by the increasing scale of artificial intelligence models in the real world, which leads to a higher demand for model inference to use pruning to reduce computation, memory, and latency\cite{frantar2023sparsegpt,ma_llm_pruner_2023,nips24_sparsellm}. 
Sparsity-based pruning improves model performance by reducing computational workload and memory usage, but also leads to accuracy loss\cite{Torsten_sparsity_in_dl}. The \textit{N:M} sparsity provides an option for balancing performance and model accuracy~\cite{mishra2021accelerating}, but introduces more complex programming and optimization challenges. The goal of our work NM-SpMM is to achieve superior performance compared to state-of-the-art methods across various sparsity levels with systematic performance analysis.


There are three major challenges in the implementation of NM-SpMM:
1) Flexibility on \textit{N:M} sparsity: 
Existing works support only a restricted set of \textit{N:M} ratios\cite{mishra2021accelerating}, or confined to specific architectures\cite{castro_venom_2023,nm_hardware_19}, or overly complex designs which hard to implementation\cite{nm_hardware_2024}. To that end, we adopt a vector-based \textit{N:M} sparsity pattern, retaining a pattern of \textit{N} vectors for every \textit{M} vectors and supports multiple vector length. This approach allows for easy integration with front-end Python APIs and also enables more selectable pruning sparsity.
2) Lack of performance analysis:
As we support more sparsity options, variation in sparsity affects the performance bottleneck from a computational bottleneck to a memory access bottleneck. To that end, we design a systematic top-down performance analysis model for \textit{N:M} sparsity. We calculate the arithmetic intensity (AI) of \textit{N:M} sparsity matrix multiplication across various matrix shapes and sparsity levels, and combine with roofline model to determine its optimization direction.
3) Low efficiency: How to select the optimal method with our performance analysis, to achieve close-to-theoretical peak performance across a wide range of sparsity levels become the final challenge.

To fix the above challenge issues, this work involves two steps: Firstly, general optimization based on dense matrix multiplication. We adapt dense GEMM optimization, such as blocking to enhance data locality and reordering to avoid bank conflict, to \textit{N:M} sparsity matrix multiplication. Secondly, we propose sparsity-aware optimization for different sparsity levels. Specially in practice, for matrix with moderate sparsity, in computational bound, we orchestration arrange instruction pipeline so that FMA instructions cover load instructions as much as possible. For high-sparsity matrix, in memory bound, we minimize the memory footprint and use load instructions to overlap FMA instructions in the pipeline.

Overall, the contributions of this paper are as follows:
\begin{itemize}
    \item NM-SpMM supports flexible \textit{N:M} ratios without dependence on specific hardware. It adopts a simple sparse pattern, retaining $N$ vectors for every $M$ vectors, facilitating its integration with algorithm research.
    \item We detail the anatomy of the \textit{N:M} sparsity computation bottleneck, revealing its transition from a computing bound to a memory bound as sparsity increases. This common characteristic aids in optimizing \textit{N:M} sparsity computation on other platforms.
    \item We propose general optimization that provide data locality through a hierarchical blocking mechanism, along with sparsity-aware optimization that reduce memory footprint and enhance pipeline design.
    \item We implement our approach effectively, with experimental results showing that our performance is 2.1x faster than nmSPARSE and 1.4x to 6.3x faster than cuBLAS's dense GEMM operations. 
\end{itemize}

The rest of this paper is organized as follows: 
Section \ref{sec:background} provides a brief introduction to \textit{N:M} sparsity and related work. 
Section \ref{sec:method} begins with a systematic analysis and introduces the optimization process step by step.
The evaluated results and performance discussion are presented in Section \ref{sec:evaluation}. 
Finally, Section \ref{sec:conclusion} concludes this work.

\section{Background}
\label{sec:background}

\begin{figure}[hbtp]
\centering
\includegraphics[width=0.9\linewidth]{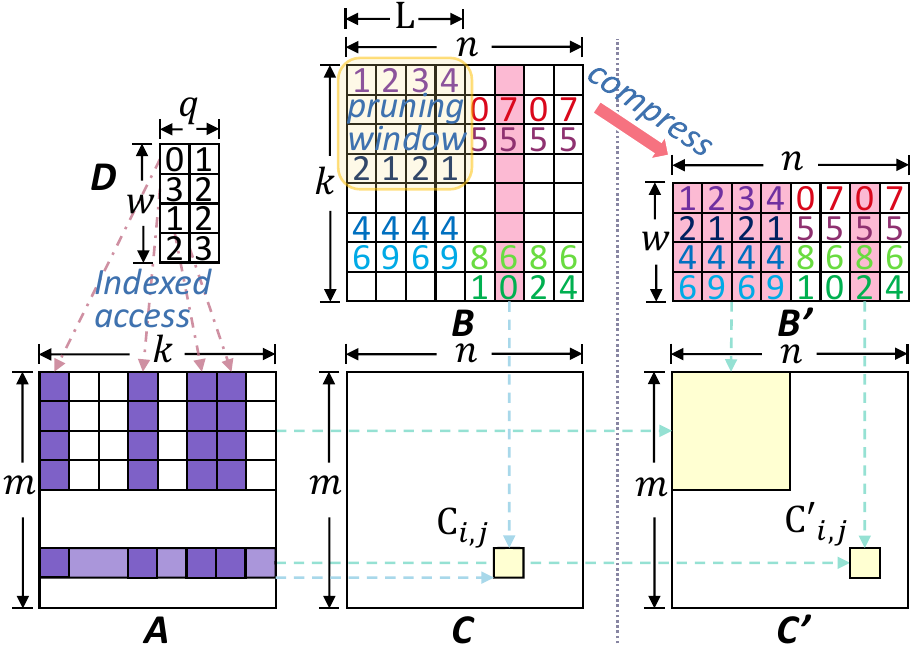}
\caption{An example demonstrates how vector-wise \textit{N:M} sparsity works, where $N=2$, $M=4$, and the vector length $L=4$.}
\label{fig:problem_desc}
\end{figure}

\subsection{\textit{N:M} Sparsity Computation}
\label{subsec:problem_desc}

To accelerate dense matrix multiplication $A = B \times C$, we utilize \textit{N:M} sparsity to prune matrix $B$, where $A$, $B$, and $C$ have dimensions $m \times k$, $k \times n$, and $m \times n$, respectively. \textit{N:M} sparsity retains $N$ units within every $M$ consecutive pruning units. This paper focuses on optimizing \textit{N:M} sparsity using vector as pruning unit. In Figure~\ref{fig:problem_desc}, we select $N$ vectors from every $M$ vector along the $k$ dimension of matrix $B$, depicted as a light yellow square frame. The selected $N$ vectors are stored in a compressed matrix $B'$ (shown in the blue box). 
An index matrix $D$ of shape $w \times q$ stores the indices of the selected $N$ vectors within each pruning window of matrix $B$, where $w = \lceil \frac{k \cdot N}{M} \rceil$ and $q = \lceil \frac{n}{L} \rceil$. We assume $k$ is divisible by $M$ and $n$ by $L$; otherwise, padding is applied, resulting in $w = \frac{k \cdot N}{M}$ and $q = \frac{n}{L}$.
Finally we get the result matrix $C'$ to approximating the original result matrix $C$ with a confusion matrix $W$: 
\begin{equation}
    C'[i][j] = \frac{M}{N} \cdot \sum_{u=0}^{w} A[i][\frac{uM}{N}+D[u][\frac{j}{L}]] \cdot B'[u][j]
\label{eq:nmsparsity}
\end{equation}

\begin{equation}
    W[i][j] = \frac{|C'[i][j] - C[i][j]|}{m \cdot n}
\label{eq:confusion}
\end{equation}
 
For convenience, we redefine the operation in equation~\ref{eq:nmsparsity} as $C = A \NMgemm(B, D)$. From the equations above, we conclude that: 1) computation operations are reduced to a ratio of $\frac{N}{M}$; 2) the same proportion of memory access on $B$ is saved, with additional data access required for matrix $D$; 3) this creates optimization opportunity for the indirect data access pattern in matrix $A$.




\subsection{Related Works}
Previous research on \textit{N:M} sparsity focuses on two categories: 1) \textit{N:M} sparsity implementation and optimization;
2) maintaining accuracy for sparse network. 

\textbf{\textit{\textit{N:M} sparsity implementation and optimization:}}
NVIDIA \cite{mishra2021accelerating} introduces Sparse Tensor Cores in the Ampere GPU architecture \cite{nvidia_A100_whitepaper}, using a 2:4 element-wise sparsity pattern to double the math throughput of dense Tensor Cores.
Castro et al. \cite{castro_venom_2023} introduces the \textit{V:N:M} format and Spatha sparse library, enabling arbitrary \textit{N:M} ratios on Sparse Tensor Cores. However, Spatha's two-stage pruning creates a pattern distinct from the \textit{N:M} sparsity studied in the algorithm community, making it incompatible with algorithms designed to preserve \textit{N:M} sparse network accuracy.
Recent studies \cite{tan_accelerating_2022, yao_balanced_2019, sc21_vector_wise_sparse_gpu_kernel, gray2017gpu, castro_probing_2023, huang_shfl-bw_2022, elsen_fast_2020,guo_accelerating_2020} show that vector-wise or block-wise sparse patterns offer significant performance benefits on modern general-purpose hardware by enhancing data reuse in L1 cache and registers.
Lin et al. \cite{lin_efficient_2023} propose nmSPARSE, which combines \textit{N:M} sparsity with vector-wise and block-wise pruning to support arbitrary \textit{N:M} ratios, eliminating the need for Sparse Tensor Cores.
But, the nmSPARSE remains sub-optimal as it does not fully exploit the locality introduced by \textit{N:M} sparsity or optimize for different sparsity levels.
Additionally, there are \textit{N:M} sparsity accelerators designed for specific network or pruning algorithm\cite{nm_hardware_19,nm_hardware_2024}.

\begin{figure*}[!h]
\centering 
 \includegraphics[width=0.9\linewidth]{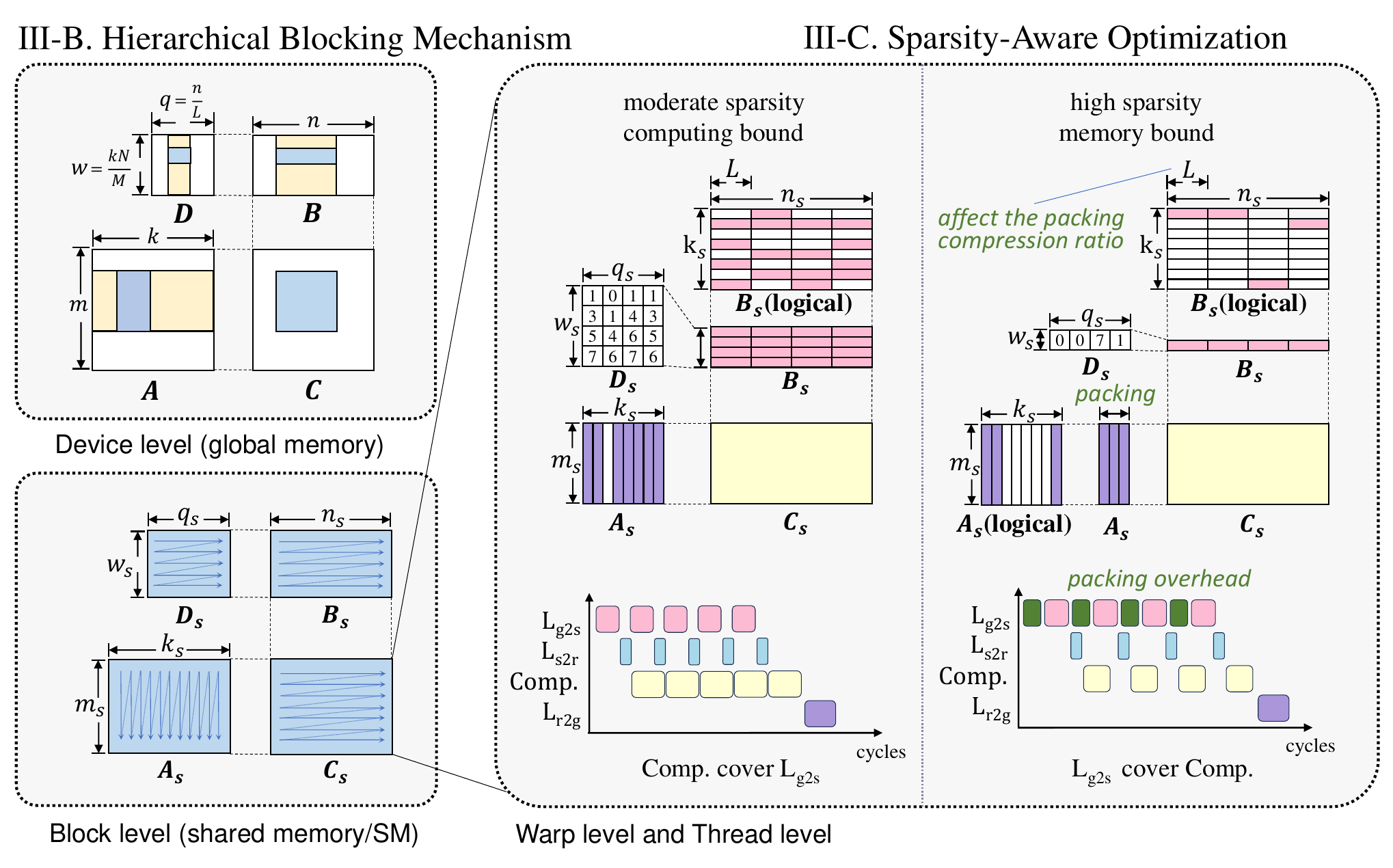}
\caption{Overview of the Workflow and Internals of NM-SpMM.}
\label{fig:overview}
\end{figure*}

\textbf{\textit{Maintaining accuracy for sparse network:}}
With the emergence of Sparse Tensor Cores, a large amount of research focuses on how to improve the accuracy of \textit{N:M} sparse networks. Our work follows the naive \textit{N:M} pattern, so we can combine it with these works to jointly promote the application of sparse networks.
NVIDIA \cite{nvidia_A100_whitepaper} introduces \textit{N:M} sparsity to accelerate inference using a standard three-step pipeline: pre-training, pruning, and fine-tuning. Pool et al. \cite{pool2021channel} enhance accuracy with channel permutation, while Sun et al. \cite{sun_dominosearch_2021, oh_attentive_2022} propose a layer-wise \textit{N:M} scheme for improved precision over uniform sparsity. Efforts to optimize \textit{N:M} training efficiency include Zhou et al.'s regularization term for learning sparsity \cite{zhou_learning_2020}, Hubara et al.'s transposable masks for faster backward passes \cite{NEURIPS2021_b0490b85}, and Zhang et al.'s Bi-Mask \cite{zhang_bi-directional_2023}. Extensions like applying \textit{N:M} to activations \cite{akiva-hochman_searching_2023}, combinatorial approaches \cite{zhang_learning_2022}, and adaptive mask learning \cite{lu_step_2023} further enhance training efficiency. Xiang et al. \cite{xiang_maxq_2024} introduce the MaxQ method for incrementally applying sparsity during training.

\section{Method}
\label{sec:method}

The whole optimization workflow of NM-SpMM is illustrated in Figure \ref{fig:overview}. Specifically, it includes a hierarchical blocking mechanism and sparsity-aware optimization. In Section \ref{subsec:analysis_for_NMGEMM}, we first provide a systematic analysis of the \textit{N:M} sparsity computation, followed by a detailed introduction to the hierarchical blocking mechanism and sparsity-aware optimization in Section \ref{subsec:Hierarchical_Blocking_Mechanism} and Section \ref{subsec:sparsity_aware_optimization}, respectively.

\subsection{Analysis of Optimization Strategies for \textit{N:M} Sparsity Computation}
\label{subsec:analysis_for_NMGEMM}

In this subsection, we first briefly introduce how we analyze the \textit{N:M} sparsity computation pattern and provide an overview of the optimization process, while outlining the key factors that influence performance.
Then, in the following subsections, we provide a detailed step-by-step explanation of our optimization approach.

The input to the \textit{N:M} sparsity computation problem consists of two matrices, $A$ and $B$, with dimensions $m\times k$ and $w\times n$, respectively. Matrix $B$ is compressed using the method shown in Figure \ref{fig:problem_desc}, as well as the $N$ and $M$ configuration is adjusted to accommodate different sparsity levels.

\begin{figure}[t]
\centering 
 \includegraphics[width=1.0\linewidth]{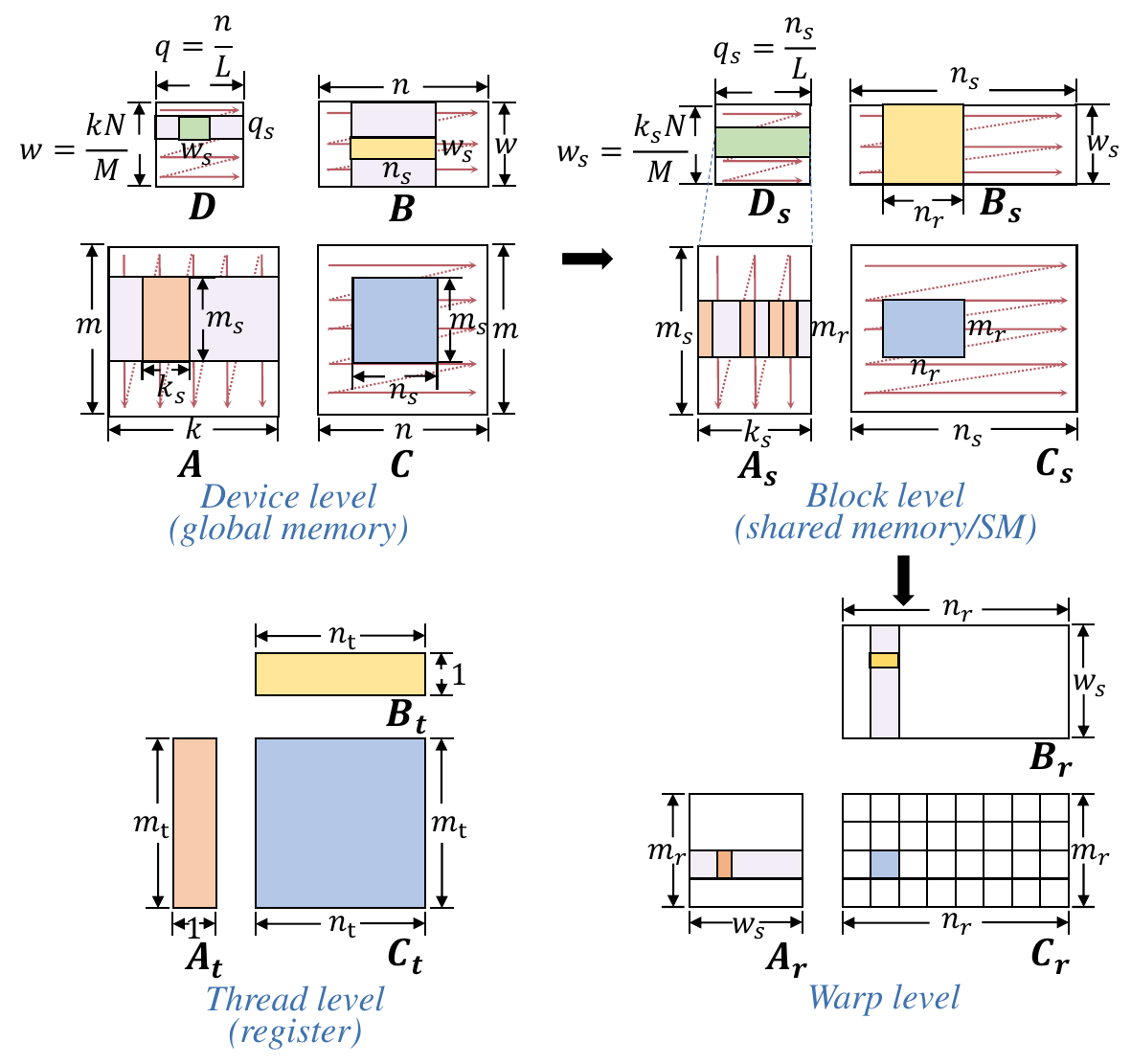}
\caption{A hierarchical pipeline of the NM-SpMM GPU implementation.}
\label{fig:multi-level_blocking}
\end{figure}

First, we use a \textbf{hierarchical blocking mechanism that adapts to the GPU's storage architecture to enhance data locality}.
Because the \textit{N:M} sparsity pattern resolves load imbalance problem in sparse matrix multiplication, the fine-grained data locality achieved by using vector as pruning unit eliminates irregular memory access issues.
Therefore, we can design the blocking method based on the optimization strategies used in dense matrix multiplication.
However, the hierarchical blocking mechanism introduces several parameter configurations, including matrix size parameters for shared memory blocking ($m_s$, $n_s$, $k_s$, $w_s$, $q_s$), register blocking parameters ($m_t$, $n_t$), and parameters affecting warp instruction scheduling ($m_r$, $n_r$), as shown in Figure \ref{fig:multi-level_blocking}.
The parameters $m_s$, $n_s$, $k_s$, and $w_s$ are constrained by the shared memory capacity on the Stream Multi-processor (SM), as indicated in Equation \ref{eq:SMsize}. 
The block parameters at the warp level and thread level ($m_t$, $n_t$, $m_r$, $n_r$) impact the performance of the inner kernel. 
It is necessary to consider the number of registers used by each thread, as well as the compute-to-memory access ratio of the inner kernel. 
This is influenced by warp-level analysis, as the GPU uses warps as the scheduling units for instructions, and shared memory access is also evaluated at the warp level.
The detailed analysis can be referenced in Section \ref{subsec:Hierarchical_Blocking_Mechanism}, and for matrices of different input size, we provide recommended parameter configurations in Table \ref{tab:tiling_parameters}. 
For handling different sparsity levels (i.e., different $N$ and $M$), we adaptively adjust the parameters based on the capacity of shared memory.

Another important point is that \textbf{varying degrees of sparsity will affect the computation pattern of \textit{N:M} sparsity}.
As shown in Figure \ref{fig:overview}, we cache the sub-matrix $A_s$ and $B_s$ of matrix $A$ and $B$ in shared memory to reduce accesses to global memory. 
The sparsity is given by \( 1 - \frac{N}{M} \) or \( 1 - \frac{w_s}{k_s} \), where \( k_s \) is typically equal to \( M \) (or sometimes a multiple of \( M \)).
In this paper, we define sparsity below 70.0\% as moderate and above 70.0\% as high. 
According to the roofline analyses and our experimental results, when the sparsity exceeds 70.0\%, the performance bottleneck shifts.
But the transition point varies depending on the arithmetic intensity of the hardware.
We select four typical levels: 50.0\%, 62.5\%, 75.0\%, and 87.5\%.
Based on the vector length $L$, $B_s$ is divided into $\frac{n_s}{L}$ pruning windows (four in Figure \ref{fig:overview}) along the row direction. 
Within each pruning window, according to the rules of matrix multiplication, the pink vector from a specific row of $B_s$ computes with the purple vector from the corresponding column in $A_s$.

One can see that the upper bound of the memory footprint of $A_s$ is $m_s \times k_s$, while the lower bound is $m_s \times w_s$. 
The memory footprint of \( A_s \) is influenced by both sparsity and vector length. Lower sparsity results in a larger memory footprint within each pruning window, while a smaller vector length \( L \) leads to more pruning windows along the row direction of \( B_s \), thereby increasing the combined memory footprint of multiple pruning windows.
Additionally, as $L$ decreases, the accuracy of the \textit{N:M} sparse network improves, while a larger $L$ facilitates load distribution within the warp and data reuse within a thread, resulting in a more efficient inner kernel.
As shown in Figure \ref{fig:overview}, in the cases of moderate sparsity and high sparsity, the memory footprint of $A_s$ accounts for $7/8$ and $3/8$ of the working set (memory address space to be accessed), respectively.
In contrast, $B_s$ is stored in a compressed format, with a fixed memory footprint of $w_s \times n_s$.
The memory footprint of $D_s$ is relatively small and can be ignored. 
The computational workload for $C_s = A_s \NMgemm (B_s, D_s)$ is $2 \times m_s \times n_s \times w_s$.
Therefore, the arithmetic intensity of \textit{N:M} sparsity can be expressed as:
\begin{equation}
\label{eq:AI_NM_sparsity}
    AI = \frac{2 \cdot m_s \cdot n_s \cdot w_s}{m_s \cdot k_s + w_s \cdot n_s + 2 \cdot m_s \cdot n_s}
\end{equation}
From Equation \ref{eq:AI_NM_sparsity}, we can conclude that as sparsity increases, the arithmetic intensity decreases. This is because, given that $w_s = k_s \cdot (1 - \text{sparsity})$, the numerator decreases with increasing sparsity, while in the denominator, only the operand $w_s \cdot n_s$ decreases proportionally. Therefore, the overall fraction decreases.
In summary, in the moderate sparsity scenario, the working set of $A_s$ is almost fully utilized, making the \textit{N:M} sparsity computation more computing bound. In contrast, in the high sparsity scenario, as sparsity increases, both the computational workload and the memory footprint of $B_s$ decrease proportionally, while the memory footprint of $A_s$ does not reduce at the same rate, resulting in a memory bound computation.

Therefore, due to the different computational characteristics in various sparsity scenarios, we propose sparsity-aware optimization: memory access design for footprint minimization and pipeline design for instruction latency hiding.
In briefly, for the computing bound problem in the moderate sparsity scenario, we load $A_s$ using a non-packing approach. 
For the memory bound problem in the high sparsity scenario, we first obtain information about the valid data needed in each block of $A_s$ through offline processing (marked as $col\_info$). 
During online computation, we then pack $A_s$ using $col\_info$ to reduce the memory footprint.

Then, we design pipelines for both scenarios to hide instruction latency. In the moderate sparsity scenario, we design the pipeline to use computation instructions to mask the instruction latency from loading data from global memory to shared memory. 
In the high sparsity scenario, due to packing requiring $col\_info$ to be loaded from global memory first, we use the instructions for loading data from global memory to shared memory to mitigate the latency of computation instructions.
Additionally, in the inner kernel, we minimize the indirect memory access impact caused by \textit{N:M} sparsity through pre-fetching. 
For more details, please refer to Section \ref{subsec:sparsity_aware_optimization}.



\subsection{Hierarchical Blocking Mechanism}
\label{subsec:Hierarchical_Blocking_Mechanism}



\lstset{
    language = C++, 
    breaklines = true, 
    breakindent = 10pt, 
    lineskip = {-1pt}, 
    basicstyle = \ttfamily\footnotesize, 
    commentstyle = {\itshape \color[rgb]{0.2,0.5,0.2}},  
    classoffset = 0, 
    keywordstyle = {\bfseries \color[rgb]{0,0,0.6}},   
    stringstyle = {\color[rgb]{0.6,0.3,0.1}},    
    frame = trbl,
    framesep = 5pt, 
    numbers = left, 
    stepnumber = 1, 
    xrightmargin = 7pt, 
    xleftmargin = 8pt, 
    numberstyle = \tiny, 
    tabsize = 3, 
    captionpos = t, 
    directivestyle = {\color{black}},
    emph = {vld1q_f32, vfmaq_f32}, 
    emphstyle = {\color[rgb]{0,0.5,0.6}},  
}

\lstset{escapeinside={<@}{@>}}

{\color{blue}
\begin{figure}[t]
\centering
\small
\begin{minipage}[c]{0.5\textwidth}
\begin{lstlisting}[caption = {Blocking algorithm for SM shared memory.},label = {list:blocking_SM},mathescape]
__global__ void <@\textbf{NM-SpMM}@>(float $A$[$m$][$k$], float $B$[$w$][$n$], float $D$[$w$][$q$], float $C$[$m$][$n$]){
  //Calculate the block size for $A_s, B_s, D_s$
  $(m_s, n_s, m_t, n_t)$ = Para_Init_Table($m$, $n$);
  $k_s$ = min($k$, $\frac{M \cdot SM\_Size}{8 \cdot (N \cdot m_s + N \cdot n_s)}$); 
  $w_s$ = $k_s \cdot N / M$, $q_s = n_s / L$; 
  //Allocate shared memory for $A_s, B_s, D_s$
  __shared__ float $A_s$[$m_s$][$k_s$], $B_s$[$w_s$][$n_s$], $D_s$[$w_s$][$q_s$];
  //Allocate accumulator registers
  float $C_t$[$m_t$][$n_t$] = {0};
  //Calculate the offset for each block
  int $b_i$ = blockIdx.y * $m_s$;
  int $b_j$ = blockIdx.x * $n_s$;
  // Main loop 
  for(int u = 0; u < $w$; u = u + $w_s$){
    //Load blocks from global to shared memory
    $A_s$=LoadTile($A$, $b_i$, u, $k$, $k_s$, $m_s$);
    $B_s$=LoadTile($B$, $b_j$, u, $n$, $w_s$, $n_s$);
    $D_s$=LoadTile($D$, $b_j$, u, $w$, $w_s$, $q_s$);
    //Synchronization after data loading
    __syncthreads();
    SMBlock($A_s$, $B_s$, $D_s$, $C_t$, $w_s$, $m_t$, $n_t$);
    //Synchronization after computation 
    __syncthreads();
  }
  //Store the results $C_t$ back to global memory
  StoreFrag($C$, $C_t$);
}

\end{lstlisting}
\end{minipage}
\end{figure}
}


\subsubsection{Blocking for SM Shared Memory}

Each Streaming Multiprocessor (SM) in GPGPU is configured with a shared memory on the chip with lower access latency compared to its global memory. Thus, to improve data locality, we first block the input matrices \( A[m][k] \), \( B[w][n] \), and \( D[w][q] \) into smaller blocks, denoted \( A_s[m_s][k_s] \), \( B_s[w_s][n_s] \), and \( D_s[w_s][q_s] \), as shown in the top-left corner of Figure \ref{fig:multi-level_blocking}. Note that $w_s =  \frac{k_s \cdot N}{M} $, $q_s =  \frac{n_s}{L}$.
Each thread block loads $A_s$, $B_s$ and $D_s$ into shared memory on each SM to compute $C_s$, this is illustrated in Figure top-right corner of \ref{fig:multi-level_blocking}.
In this blocking process, the block size of these three matrices is \textbf{limited by the memory size of the SM shared memory.} Thus the block size parameters must satisfy the equation below:
\begin{equation}
\label{eq:SMsize}
    4 \cdot (k_s \cdot m_s + w_s \cdot n_s + w_s \cdot q_s) \leq SM\_Size * 0.5
\end{equation}
Here $SM\_Size$ denotes the shared memory size on each SM. We keep half of its shared memory for buffering and other temporary variable. 
Since the index matrix $D$ only needs to provide the position of each retained vector within the pruning window, each element requires only $\log_2{M}$ bits.
Thus we ignore the shared memory size used by $D_s[w_s][q_s]$ to simplify equation ~\ref{eq:SMsize}, then we can update equation~\ref{eq:SMsize} to be:
\begin{equation}
\label{eq:SMsize2}
    8 \cdot k_s (m_s + \frac{N \cdot n_s}{M}) \leq SM\_Size 
\end{equation}

We also need to \textbf{maximize the block-level arithmetic intensity}, as shown in Equation \ref{eq:AI_NM_sparsity}.
The actual value for $m_s$ and $n_s$ are the user-defined parameter settings. Different
settings are selected based on the size of the input matrix size, some recommended configurations for $m_s$ and $n_s$ are given in Table \ref{tab:tiling_parameters} to accommodate small, medium and large matrices. \textbf{To avoid bank conflict in shared memory access, $m_s$ and $n_s$ are set as multiples of 32.} Once the values of \( m_s \) and \( n_s \) are determined, the maximum \( k_s \) can be directly calculated using Equation~\ref{eq:SMsize}, as \textbf{a larger \( k_s \) ensures sufficient compute instructions in the pipeline of inner kernel.}

\begin{table}[htbp]
    \centering
    \begin{tabularx}{\linewidth}{X X X  | X X  | X X}
        \toprule 
             & $m_s$ & $n_s$ & $m_r$ & $n_r$ & $m_t$ & $n_t$ \\ 
        \midrule
        small  & 32 & 32  & 16 & 32  & 4  & 4  \\ 
        medium & 32 & 64  & 32 & 32  & 8  & 4  \\ 
        large   & 64 & 128 & 64 & 32  & 8  & 8  \\ 
        \bottomrule 
    \end{tabularx}
    \caption{Recommended parameter configurations for SM shared memory blocking, warp-level tilling and thread tilling.} 
    \label{tab:tiling_parameters}
\end{table}

We demonstrate this blocking process in Listing \ref{list:blocking_SM}. In line 3, we first initiate the blocking parameters $m_s, n_s, m_t, n_t$ using the configurations given by Table \ref{tab:tiling_parameters}. $k_s$ are calculated with the formula given by equation~\ref{eq:SMsize} in line 4.  Then line 7 declares and allocates the SM shared memory space for $A_s$, $B_s$, and $D_s$. We also assign a temporary matrix $C_t[m_t][n_t]$ as a result accumulator and initialize it with zeros. The offset for each block is calculated by lines 11 to 12. We loop over the $k_s$ and $w_s$ dimension with a step $k_s$ and $w_s$ respectively between lines 14 to 24. In each iteration $A_s$, $B_s$, and $D_s$ are loaded into shared memory, then the function \textit{SMBlock} is routed to calculate the sub-problem. Finally, the accumulated result is saved back to global memory in line 26.  

The following subsection introduce how we implement the function \textit{SMBlock}, and to fully utilize the registers, we tiling each block again to fit each tile into the registers and dispatch the workload to each thread hosting one FP32 ALU.

\lstset{
    language = C++, 
    breaklines = true, 
    breakindent = 10pt, 
    lineskip = {-1pt}, 
    basicstyle = \ttfamily\footnotesize, 
    commentstyle = {\itshape \color[rgb]{0.2,0.5,0.2}},  
    classoffset = 0, 
    keywordstyle = {\bfseries \color[rgb]{0,0,0.6}},   
    stringstyle = {\color[rgb]{0.6,0.3,0.1}},    
    frame = trbl,
    framesep = 5pt, 
    numbers = left, 
    stepnumber = 1, 
    xrightmargin = 7pt, 
    xleftmargin = 8pt, 
    numberstyle = \tiny, 
    tabsize = 3, 
    captionpos = t, 
    directivestyle = {\color{black}},
    emph = {vld1q_f32, vfmaq_f32}, 
    emphstyle = {\color[rgb]{0,0.5,0.6}},  
}
\lstset{escapeinside={<@}{@>}}

{\color{blue}
\begin{figure}[t]
\centering
\small
\begin{minipage}[c]{0.5\textwidth}
\begin{lstlisting}[caption = {Warp tiling algorithm with thread inner-kernel optimization.},label = {list:Warp_tiling},mathescape]
__device__ void <@\textbf{SMBlock}@>(float $A_s$[$k_s$][$m_s$], float $B_s$[$w_s$][$n_s$], float $D_s$[$w_s$][$q_s$], float $C_t$[$m_t$][$n_t$], int $w_s$, int $m_t$, int $n_t$){
  // Allocate register buffer
  float $A_t$[$m_t$], $B_t$[$n_t$];      
  // Calculate the offset of Warp tile
  int $t_i$, $t_j$;
  ThreadIndexing($t_i$, $t_j$, $m_t$, $n_t$);
  // SM block loop
  for(int p = 0; p < $w_s$; p++){
    // Load $A_t$,$B_t$ from shared memory to register      
    $A_t$=LoadFragByIdx($A_s$, $t_i$, p, $D_s$);
    $B_t$=LoadFrag($B_s$, $t_j$, p);
    // Calculate the outer product of $A_t$ and $B_t$
    InnerKernel($A_t$, $B_t$, $C_t$, $m_t$, $n_t$);
  }
}

__device__ void <@\textbf{ThreadIndexing}@>(int& $t_i$, int& $t_j$, int $m_t$, int $n_t$){
  int tid = threadIdx.y*blockDim.x + threadIdx.x;
  int warp_id = tid / warpSize; 
  int lane_id = tid % warpSize; 
  // Arrange the threads in a warp into a 4x8 grid
  $t_i$ = warp_id/2 * $m_t$ * 4 + lane_id/8 * $m_t$;            
  $t_j$ = warp_id%2 * $n_t$ * 8 + lane_id%8 * $n_t$;       
}

__device__ void <@\textbf{InnerKernel}@>(float $A_t$[$m_t$], float $B_t$[$n_t$], float $C_t$[$m_t$][$n_t$], int $m_t$, int $n_t$){
  // Calculate the outer product of $A_t$ and $B_t$
  for(int i = 0; i < $m_t$; i++){
    for(int j = 0; j < $n_t$; j++){
      $C_t$[i][j] += $A_t$[i] * $B_t$[j]; 
    }  
  }
}

\end{lstlisting}
\end{minipage}
\end{figure}
}


\subsubsection{Warp Tiling for Thread Inner Kernel Optimization}
Blocks from the above subsection are sent to each SM for calculation, a further warp tiling method is needed to dispatch the workload to multiple warps (or a cluster of threads). NVIDIA GPUs are configured with hundreds of SMs. Each SM utilizes a single-instruction multiple-thread (SIMT) programming model and can simultaneously launch up to 64 warps on the A100. Each warp consists of 32 threads, and the thread arrangement within a warp is performance sensitive. A warp with 32 threads can be arranged in a grid of $x \times y$, for example $1\times 32$, $2\times 16$, $4\times 8$, $8\times 4$, etc. The more square grid $4\times 8$ or $8\times 4$ are better in most cases. Figure \ref{fig:multi-level_blocking} (bottom left) illustrates an example with a $4\times 8$ grid, where the 8 threads in each row access different banks of $B_s$ and broadcast to the 4 threads in the corresponding column. 32 threads in a wrap are scheduled out-of-order in parallel to one stream processor (SP), and only 16 threads can be executed simultaneously, for each SP has only 16 FP32 cores. 

The arithmetic intensity of the inner kernel executed by each thread is crucial for achieving high computation efficiency. Since each thread runs on a physical FP32 core and is limited to 255 registers, we aim to \textbf{maximize the computing-to-memory-access ratio (CMAR) of the inner kernel} to optimize performance. We first extract compressed columns from $A_s$ with the indices provided by $D_s$ to form a new matrix $A_r$ during warp tiling. Then the innermost computation for the thread transforms into a general matrix multiplication (GEMM). To maximize CMAR,  we will increase the FP32 FMA instructions to LDS (load from shared memory) instructions for each thread's inner kernel while ensuring that there are no bank conflict. Therefore, a thread tilling is applied with a thread tile of size $m_t \times n_t$, the number of FMA instructions in the thread block is $m_t \times n_t$, and the number of LDS instructions is proportional to the sum of $m_t$ and $n_t$ (with the proportionality constant $\alpha$, which depends on the LDS access width, such as $\alpha=4$ in LDS.32, $\alpha=2$ in LDS.64,  and $\alpha=1$ in LDS.128). Thread block size should satisfy $(m_t + n_t + m_t \cdot n_t) \leq 255$, where 255 refers to the maximum number of registers available per thread. Because \textbf{using too many registers per thread reduces parallelism}, which is referred to as occupancy in NVIDIA GPUs. Under this constraint, the CMAR is maximized:
\begin{equation}
\begin{aligned}
    CMAR = \frac{1}{\alpha} \cdot \frac{m_t \cdot n_t}{m_t + n_t} \\
\end{aligned}
\end{equation}
One can see that the larger $m_t$ and $n_t$ are, the higher the CMAR is. On A100, $m_t$ and $n_t$ are typically set to $8 \times 8$ or $8 \times 16$. For small matrices, using a smaller thread tile size, such as $8 \times 4$ or $4 \times 4$, reduces the resource usage per thread. This allows more threads to be launched simultaneously, thereby increasing occupancy and improving computational efficiency.

The Warp tiling algorithm and thread inner kernel optimization are presented in Listing~\ref{list:Warp_tiling} and Figure \ref{fig:multi-level_blocking} (bottom left). We first allocate registers for $A_t$ and $B_t$ to load data for thread inner kernel in line 3. Then the thread index is calculated by the function \textit{ThreadIndexing} in line 6. The implementation of  the function \textit{ThreadIndexing} is presented from line 17 to line 24, and here we provide an example of its implementation using grid $4 \times 8$. After that, each thread loads $A_t$, $B_t$ into registers for computation with thread \textit{InnerKernel} function. 
Its detailed implementation is presented from line 26 to line 33. Finally the results $C_t$ are accumulated in the registers before being written back to $C$ in global memory.


\subsection{Sparsity-Aware Optimization}
\label{subsec:sparsity_aware_optimization}

\subsubsection{Memory Access Design for Footprint Minimization}
In Section \ref{subsec:analysis_for_NMGEMM}, we explore how different levels of sparsity impact the computation pattern, emphasizing the distinctions between moderate and high sparsity. As a result, we propose packing strategy to address the high sparsity scenario and non-packing strategy to resolve the moderate sparsity scenario. 

\begin{figure}[t]
\centering 
\includegraphics[width=1.0\linewidth]{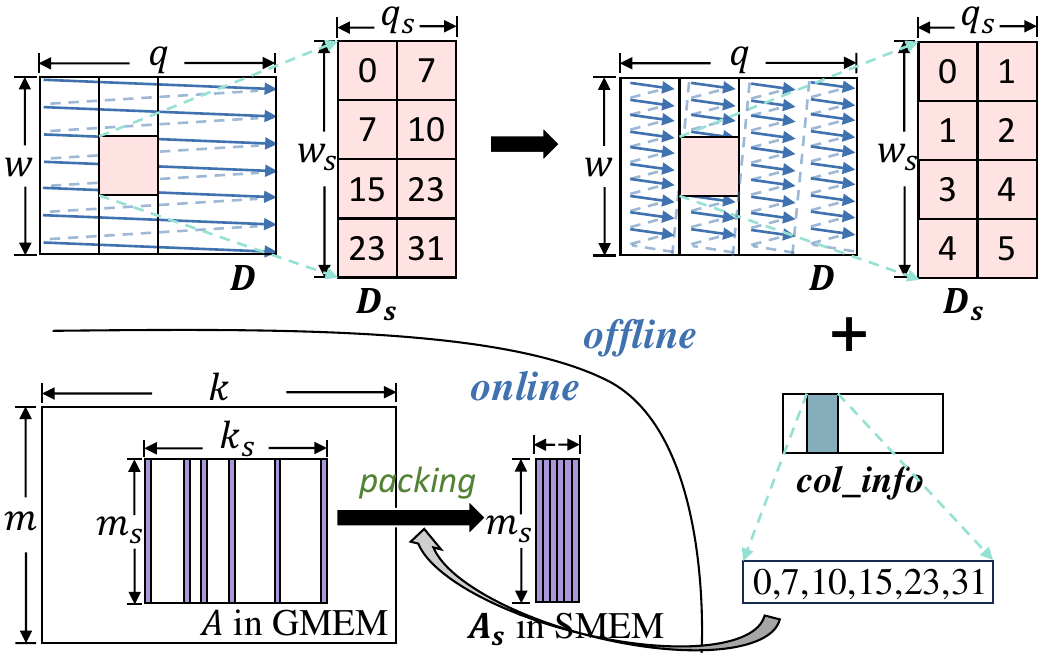}  
\caption{Offline pre-processing of the index matrix in high sparsity scenarios: obtaining $col\_info$, rearranging indices, and changing the data layout. During computation, online packing of $A_s$ reduces memory footprint and enhances arithmetic intensity.}
\label{fig:packing}
\end{figure}

We design a packing approach to minimize the memory footprint of matrix $A_s$ and eliminate redundant reads of matrix $A$ in high sparsity scenarios.
For example, packing can reduce the global memory access of $A_s$ to $ \frac{n_s \cdot N}{L \cdot M} $, where $ \frac{n_s}{L} $ represents the number of pruning window per row in $B_s$. When the pattern of each pruning window is identical, the memory access minimize to $ \frac{N}{M} $.
Specifically, we conduct offline pre-processing to accomplish three tasks.
First, we identify the necessary columns in $A_s$, referred to as $col\_info$.
Second, we reorder the index matrix $D$ to establish the mapping for \textit{N:M} sparsity. 
Third, we transform the data layout of matrix $D$ to reduce the number of global memory transactions.
The offline pre-processing procedure is shown in Figure \ref{fig:packing} and Listing \ref{list:reduce_footprint} lines 2-6. 
During computation, we online pack $A_s$ by $col\_info$ to reduce memory footprint and increase arithmetic intensity.
Before loading $A_s$, we first load $col\_info$ from global memory, which increases latency. In Section \ref{subsubsec:pipeline_design}, we design a refined pipeline to mask this latency. The $col\_info$ introduces an additional 1\% to 10\% GPU memory overhead, calculated as $\frac{k_s}{w_s \cdot q_s}$\% of $D_s$, which is negligible in practice.

The non-packing strategy directly loads the entire working set of \( A_s \) into shared memory for moderate sparsity scenarios in an ostrich-style approach.
For moderate sparsity scenarios, the valid data in \( A_s \) occupies the majority of its working set.
Therefore, non-packing does not cause redundant reads, and skipping packing avoids the overhead of pre-processing and reading $col\_info$.
There are two main reasons for the high proportion of effective data in $A_s$ in the scenario of moderate sparsity:
a) When sparsity is moderate, the proportion of valid data within a single pruning window is relatively high.
b) For both \textit{N:M} sparse network accuracy and computational efficiency considerations, there are multiple pruning windows in the row direction of $B_s$, each often with a different pattern, as shown in Figure \ref{fig:overview}.
As shown in Listing \ref{list:reduce_footprint} lines 13-21, for high sparsity scenarios, we use a packing strategy to avoid redundant global memory accesses. For moderate sparsity scenarios, we load the data directly without packing for efficiency.

\lstset{
    language = C++, 
    breaklines = true, 
    breakindent = 10pt, 
    lineskip = {-1pt}, 
    basicstyle = \ttfamily\footnotesize, 
    commentstyle = {\itshape \color[rgb]{0.2,0.5,0.2}},  
    classoffset = 0, 
    keywordstyle = {\bfseries \color[rgb]{0,0,0.6}},   
    stringstyle = {\color[rgb]{0.6,0.3,0.1}},    
    frame = trbl,
    framesep = 5pt, 
    numbers = left, 
    stepnumber = 1, 
    xrightmargin = 7pt, 
    xleftmargin = 8pt, 
    numberstyle = \tiny, 
    tabsize = 3, 
    captionpos = t, 
    directivestyle = {\color{black}},
    emph = {vld1q_f32, vfmaq_f32}, 
    emphstyle = {\color[rgb]{0,0.5,0.6}},  
}
\lstset{escapeinside={<@}{@>}}

{\color{blue}
\begin{figure}[t]
\centering
\small
\begin{minipage}[c]{0.5\textwidth}
\begin{lstlisting}[caption = {NM-SpMM with sparsity-aware optimization to reduce memory footprint.},label = {list:reduce_footprint},mathescape]
// Offline init when sparsity is high
void <@\textbf{PreProcessing}@>(float $A$[$m$][$k$], float $B$[$w$][$n$], float $D$[$w$][$q$]){
  $col\_info$ = queryColInfo($D$, $w_s$, $q_s$);       
  $D$ = reoderingIdx($D$, $w_s$, $q_s$, $col\_info$);
  $D$ = transformLayout($D$, $w_s$, $q_s$);                                
}    
__global__ void <@\textbf{NM-SpMM}@>(float $A$[$m$][$k$],float $B$[$w$][$n$],float $D$[$w$][$q$],int* $col\_info$,float $C$[$m$][$n$]){
  <@\textbf{...}@>
  __shared__ int $sh\_col\_info$[$k_s$];  
  // Main loop 
  for(int u = 0; u < $w$; u = u + $w_s$){
    // Load SM Block from global memory to shared memory
    if(sparsity > threshold){
      // packing load
      $sh\_col\_info$ = loadColInfo($col\_info$, $b_j$, u);
      __syncthreads();
      $A_s$ = LoadTileByColInfo($A$, $b_i$, u, $sh\_col\_info$);
    } else {
      // non-packing load
      $A_s$ = LoadTile($A$, $b_i$, u);
    }                                     
    $B_s$ = LoadTile($B$, $b_j$, u);                                        
    $D_s$ = LoadTile($D$, $b_j$, u);   
    // Synchronization for data loading
    __syncthreads();
    SMBlock($A_s$, $B_s$, $D_s$, $C_t$, $w_s$, $m_t$, $n_t$);
    // Synchronization for computation completion
    __syncthreads();
  }
  // Write result to global memory
  StoreFrag($C$, $C_t$, $b_i$ + $t_i$, $b_j$ + $t_j$);
}

\end{lstlisting}
\end{minipage}
\end{figure}
}


\subsubsection{Pipeline Design for Instruction Latency Hiding}
\label{subsubsec:pipeline_design}

In Section \ref{subsec:Hierarchical_Blocking_Mechanism}, we leverage data locality provided by the hierarchical blocking mechanism to accelerate computation. In Section \ref{subsec:sparsity_aware_optimization}, we perform sparsity-aware memory access optimization at the thread block level.
However, in Listing \ref{list:reduce_footprint}, there are still load-compute and load-load dependencies that prevent the compute units from being fully utilized.
Specifically, load-compute dependencies occur between loading $A_s$, $B_s$, and $D_s$, or $A_t$ and $B_t$ and the subsequent computation.
Load-load dependencies occur between loading $col\_info$ and loading $A_s$, as well as between reading indices from $D_s$ and loading $A_t$.

\begin{figure}[h]
\centering 
 \includegraphics[width=1.0\linewidth]{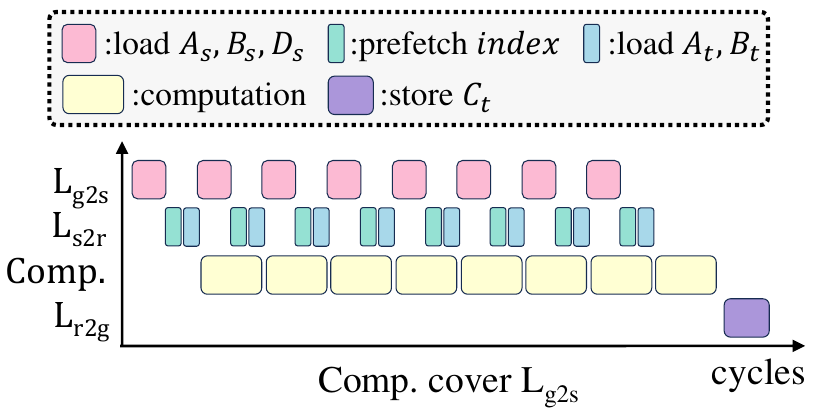}
\caption{Pipeline of NM-SpMM for moderate sparsity scenario: utilizing computation instructions to mask latency of load instructions from global memory to shared memory.}
\label{fig:pipeline_moderate}
\end{figure}

\begin{figure}[h]
\centering 
 \includegraphics[width=1.0\linewidth]{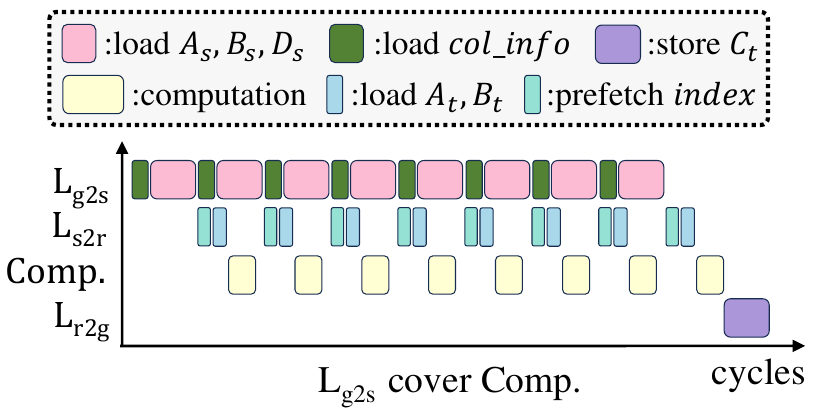}
\caption{Pipeline of NM-SpMM for high sparsity scenario: employing instructions for loading data from global memory into shared memory to hide latency of computation instructions.}
\label{fig:pipeline_high}
\end{figure}

We design pipelines for high sparsity and moderate sparsity to break these two dependencies, thereby hiding instruction latency and enhancing instruction-level parallelism while improving hardware utilization.
In the hierarchical blocking mechanism, a large number of registers are used to accumulate $C_t$, resulting in lower occupancy and limited potential for increasing thread-level parallelism.
The key to using pipelines to hide instruction latency is identifying which set of instructions can mask the latency of another set. 
Based on the analysis in Section \ref{subsec:analysis_for_NMGEMM}, in the moderate sparsity scenario, we use computation instructions to mask the latency of data transfer instructions from global memory to shared memory, as shown in Figure \ref{fig:pipeline_moderate}. 
In contrast, in the high sparsity scenario, due to the overhead from packing and the inability to proportionally reduce the memory footprint of $A_s$, we utilize instructions that load data from global memory to shared memory to hide the latency of computation instructions, as illustrated in \ref{fig:pipeline_high}.

Specifically, we utilize double buffering to achieve overlap between different types of instructions.
Therefore, in Listing \ref{list:nm-spmm}, we allocate double the buffer space (line 4,31), computing on data from one buffer while writing the next round data to the other buffer.
In Listing \ref{list:nm-spmm}, lines 14 to 24 contain the main loop, where line 17 loads the next round of the SM block, line 19 performs computations based on the current SM block, and line 20 waits for data loading to complete.
Outside the main loop, line 10 is responsible for loading the first SM block, and line 26 computes the final SM block.
The \textit{SMBlock} function follows a similar logic: lines 39-40 read data for the next warp tiling, line 42 performs computations on the current warp tiling, and lines 34-35 and 45 handle the processing of the first and final warp tiling, respectively.
Unlike previously, there are no explicit asynchronous instructions required in this case. By issuing shared memory access instructions prior to the computation instructions, we enable the overlap of data transfer from shared memory to registers with computation, as illustrated by the blue and yellow rectangles in Figure \ref{fig:pipeline_moderate}.
To avoid excessive shared memory access instructions in the inner kernel, we pre-fetch the indices required by each thread from shared memory ($D_s$) into registers, as shown in Listing \ref{list:nm-spmm} lines 12 and 23. This enhances the computation-to-memory access ratio in the inner kernel.


\lstset{
    language = C++, 
    breaklines = true, 
    breakindent = 10pt, 
    lineskip = {-1pt}, 
    basicstyle = \ttfamily\footnotesize, 
    commentstyle = {\itshape \color[rgb]{0.2,0.5,0.2}},  
    classoffset = 0, 
    keywordstyle = {\bfseries \color[rgb]{0,0,0.6}},   
    stringstyle = {\color[rgb]{0.6,0.3,0.1}},    
    frame = trbl,
    framesep = 5pt, 
    numbers = left, 
    stepnumber = 1, 
    xrightmargin = 7pt, 
    xleftmargin = 8pt, 
    numberstyle = \tiny, 
    tabsize = 3, 
    captionpos = t, 
    directivestyle = {\color{black}},
    emph = {vld1q_f32, vfmaq_f32}, 
    emphstyle = {\color[rgb]{0,0.5,0.6}},  
}
\lstset{escapeinside={<@}{@>}}

{\color{blue}
\begin{figure}[t]
\centering
\small
\begin{minipage}[c]{0.5\textwidth}
\begin{lstlisting}[caption = {NM-SpMM with sparsity-aware optimization to hide instruction latency.},label = {list:nm-spmm},mathescape]
__global__ void <@\textbf{NM-SpMM}@>(float $A$[$m$][$k$],float $B$[$w$][$n$],float $D$[$w$][$q$],int *$col\_info$,float $C$[$m$][$n$]){
  <@\textbf{...}@>
  // Allocate double buffer
  __shared__ float $A_s$[2][$k_s$][$m_s$], $B_s$[2][$w_s$][$n_s$], $D_s$[2][$w_s$][$q_s$], col_info[2][$k_s$];
  float $C_t$[$m_t$][$n_t$]={0}; 
  // Buffer for prefetching indices 
  int idx[$w_s$];
  int cur = 1, nex = 0; 
  // Load the first SM Block
  LoadTile($A$, $B$, $D$, $A_s$, $B_s$, $D_s$, $b_i$, $b_j$, 0, cur);
  // Prefetch indices to register
  idx = prefetch($D_s$, 0, $b_j$);
  // Main loop 
  for(int u = $w_s$; u < $w$; u = u + $w_s$){
    cur = cur ^ 1, nex = nex ^ 1;
    // Load the next SM block async
    LoadTileAsync($A$, $B$, $D$, $A_s$, $B_s$, $D_s$, $b_i$, $b_j$, u, nex);
    // Compute on the cur SM block
    SMBlock($A_s$[cur],$B_s$[cur],idx,$C_t$,$w_s$,$m_t$,$n_t$);
    WaitAsyncCopy();
    __syncthreads();
    // Prefetch indices to register
    idx = prefetch($D_s$, u, $b_j$);
  }
  // Compute on the last SM block
  SMBlock($A_s$[cur^1],$B_s$[cur^1],idx,$C_t$,$w_s$,$m_t$,$n_t$);
  // Write result to global memory
  StoreFrag($C$, $C_t$, $b_i$ + $t_i$, $b_j$ + $t_j$);
}
__device__ void <@\textbf{SMBlock}@>(float $A_s$[$k_s$][$m_s$], float $B_s$[$w_s$][$n_s$], int idx[$w_s$], float $C_t$[$m_t$][$n_t$], int $w_s$, int $m_t$, int $n_t$){
  float $A_t$[2][$m_t$],$B_t$[2][$n_t$]; int $t_i$, $t_j$;
  ThreadIndexing($t_i$, $t_j$);
  // Load the first warp tiling
  $A_t$[0] = LoadFragByIdxInReg($A_s$,$t_i$,0,idx);
  $B_t$[0] = LoadFrag($B_s$, $t_j$, 0);
  // SM block loop
  for(int p = 0; p < $w_s$ - 1; p = p + 1){
    // Load the next warp tiling
    $A_t$[(p+1)%2]=LoadFragByIdxInReg($A_s$,$t_i$,p+1,idx);
    $B_t$[(p+1)%2] = LoadFrag($B_s$, $t_j$, p+1);     
    // Compute on the cur warp tiling
    InnerKernel($A_t$[p%2], $B_t$[p%2], $C_t$, $m_t$, $n_t$);
  }
  // Compute on the last warp tiling
  InnerKernel($A_t$[1], $B_t$[1], $C_t$, $m_t$, $n_t$);
}

\end{lstlisting}
\end{minipage}
\end{figure}
}


\section{Performance Evaluation}
\label{sec:evaluation}

\subsection{Datasets and Evaluation Environments}

\textbf{Our dataset} consists of 100 data points. These data points are extracted from linear layers in Llama\cite{touvron2023llama} models. In detail, the input sequence $m$ ranges from $2^8$ to $2^{12}$, yielding five distinct values. Each value is associated with 20 data points, where the tuples $(n, k)$ are extracted from the Llama model. Then there are 100 combinations of $(m, n, k)$ in our dataset. In addition, to validate the kernels we design for small, medium, and large matrices (see Table \ref{tab:tiling_parameters}), we select several small, medium, and large input matrices as test cases, as shown in Table \ref{tab:matrix_sizes}.

\textbf{Our evaluation environments} include three GPU cards: one NVIDIA A100 80GB PCIe, one NVIDIA RTX 3090, and one NVIDIA RTX 4090. Some key parameters of these three GPUs are shown in Table \ref{tab:gpu-comparison}. The operating system on our test is CentOS 7.9 with CUDA 12.2. We implement our work using C++ and CUDA, and name it \textbf{NM-SpMM}. NM-SpMM is compared with all state-of-the-art dense and sparse libraries, including cuBLAS (a vendor-specific library for dense matrix computation by NVIDIA), nmSPARSE \cite{lin_efficient_2023} (a state-of-the-art library of general \textit{N:M} sparsity implementation), and Sputnik \cite{gale2020sparse} (an excellent library of sparse linear algebra kernels for deep learning). Without loss of generality, four typical sparsity ratios—50.0\%, 62.5\%, 75.0\%, and 87.5\%—are used in our benchmark evaluations.

\begin{table}[t]
    \centering
    \begin{tabular}{ccccccc}
        \toprule
        & \multicolumn{2}{c}{small} & \multicolumn{2}{c}{medium} & \multicolumn{2}{c}{large} \\
        \cmidrule(lr){2-3} \cmidrule(lr){4-5} \cmidrule(lr){6-7}
        label & A & B & C & D & E & F \\
        m & 512 & 512  & 512  & 1024 & 2048 & 4096 \\
        n & 512 & 1024 & 2048 & 2048 & 4096 & 4096 \\
        k & 512 & 1024 & 2048 & 2048 & 4096 & 4096 \\
        \bottomrule
    \end{tabular}
    \caption{Examples of small, medium, and large matrices for evaluating kernels with different blocking parameters in Table \ref{tab:tiling_parameters}.}
    \label{tab:matrix_sizes}
\end{table}

\begin{table}[t]
    \centering
    {\fontsize{6}{8}\selectfont 
    \renewcommand{\arraystretch}{1.5} 
    \begin{tabular}{|p{4.0cm}|p{1cm}|p{1cm}|p{1cm}|} 
        \hline
        \rowcolor{gray!20} 
        Hardware Metric & A100 80G & RTX 3090 & RTX 4090 \\
        \hline
        Boost Clock (MHz) & 1410 & 1695 & 2520 \\
        \hline
        Peak FP32 TFLOPS & 19.5 & 35.6 & 82.6 \\
        \hline
        Number of SMs & 108 & 82 & 128 \\
        \hline
        Register File Size / SM (KB) & 256 & 256 & 256 \\
        \hline
        FP32 Cores / SM & 64 & 128 & 128 \\
        \hline
        FP32 FLOPs / clock / SM & \textbf{128} & \textbf{256} & \textbf{256} \\
        \hline
        L1 Data Cache / Shared Memory / SM (KB) & 192 & 128 & 128 \\
        \hline
        L2 Cache Size (MB) & 40 & 6 & 72 \\
        \hline
        Global Memory (DRAM) Size (GB) & 80 & 24 & 24 \\
        \hline
        DRAM Bandwidth (GB/s) & \textbf{1935} & \textbf{936} & \textbf{1008} \\
        \hline
    \end{tabular}
    }
    \caption{Hardware metrics comparison between A100 80G PCIe, RTX 3090, and RTX 4090 GPUs.}
    \label{tab:gpu-comparison}
\end{table}

\subsection{Evaluation on Step-wise Optimizations}

The proposed step-wise optimizations in Section \ref{sec:method} are first evaluated for effectiveness. We name three versions for this experiment:
V1 relates to the hierarchical blocking mechanism found in Listing \ref{list:blocking_SM} and Listing \ref{list:Warp_tiling}.
In contrast, V2 focuses on the sparsity-aware memory footprint optimization presented in Listing \ref{list:reduce_footprint}.
Lastly, V3 addresses the sparsity-aware instruction latency hiding optimization detailed in Listing \ref{list:nm-spmm}.
Each version incrementally builds on the previous one: V2 includes V1's optimizations, and V3 encompasses both V1 and V2's improvements.

\begin{figure}[t]
\centering 
 \includegraphics[width=1.0\linewidth]{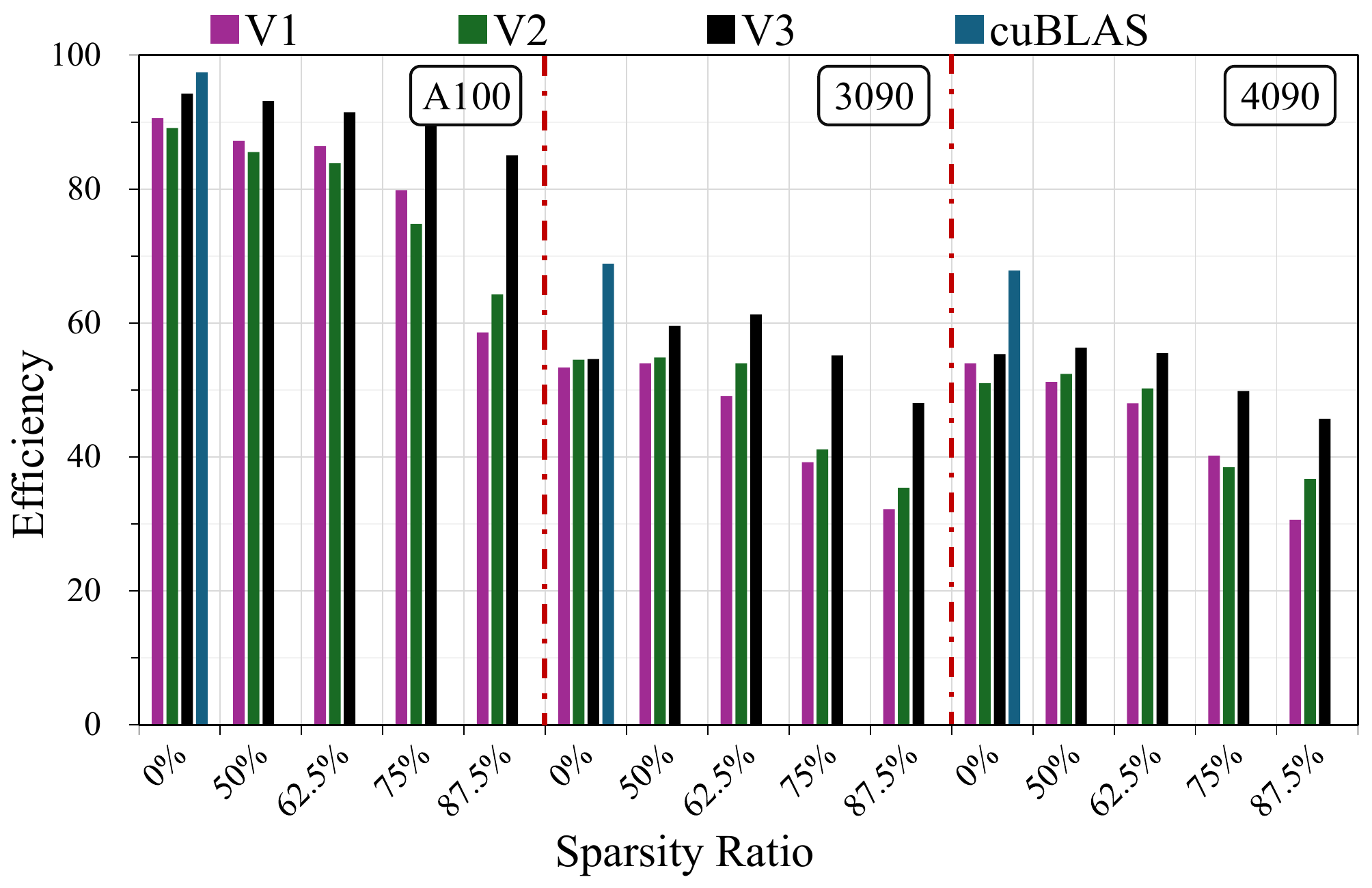}  
\caption{Step-wise optimization evaluation of NM-SpMM on A100 with input matrix shape $m = n = k = 4096$.} 
\label{fig:step_wise}
\end{figure}

In this experiment, square matrices with dimensions $m = n = k = 4096$ are used as our input data.
We select four commonly used sparsity levels in deep learning: 50.0\%, 62.5\%, 75.0\%, and 87.5\%. 
In addition, there is a sparsity level of 0.0\%, where our code sets $M = N = 32$, and cuBLAS performs dense matrix multiplication operations.
As illustrated in Figure \ref{fig:step_wise}, we evaluate the efficiency of V1, V2, and V3 on three different GPUs: A100, 3090, and 4090. 

When the sparsity is 0.0\%, cuBLAS performs dense matrix multiplication, while NM-SpMM handles \textit{N:M} sparsity computation with both $N$ and $M$ set to 32.
Through our optimizations, on A100, the computation of \textit{N:M} sparsity is comparable in efficiency to the cuBLAS dense matrix multiplication, indicating that the overhead of indirect memory access is well hidden. 
However, on the 3090 and 4090, the floating-point performance significantly outpaces memory bandwidth (see Table \ref{tab:gpu-comparison}), making it challenging to mask the overhead of indirect memory access. 

When the sparsity is 50.0\% or 62.5\%, \textit{N:M} sparsity computations tend computing bound. The hierarchical blocking mechanism implemented in V1 provides good data locality, resulting in strong performance for V1, while subsequent versions show only minor improvements.

When the sparsity is 75\% or 87.5\%, \textit{N:M} sparsity computations become memory bound. The optimizations in V2 that reduce memory footprint, along with the pipeline optimizations in V3 that hide instruction latency, significantly enhance performance.
For GPUs like the 3090 and 4090, where memory bandwidth is somewhat constrained relative to computational power, these optimizations targeting memory bound scenarios yield even more significant improvements.

\subsection{Evaluation of Kernels with Different Blocking Parameters}

\begin{figure}[t]
\centering 
 \includegraphics[width=1.0\linewidth]{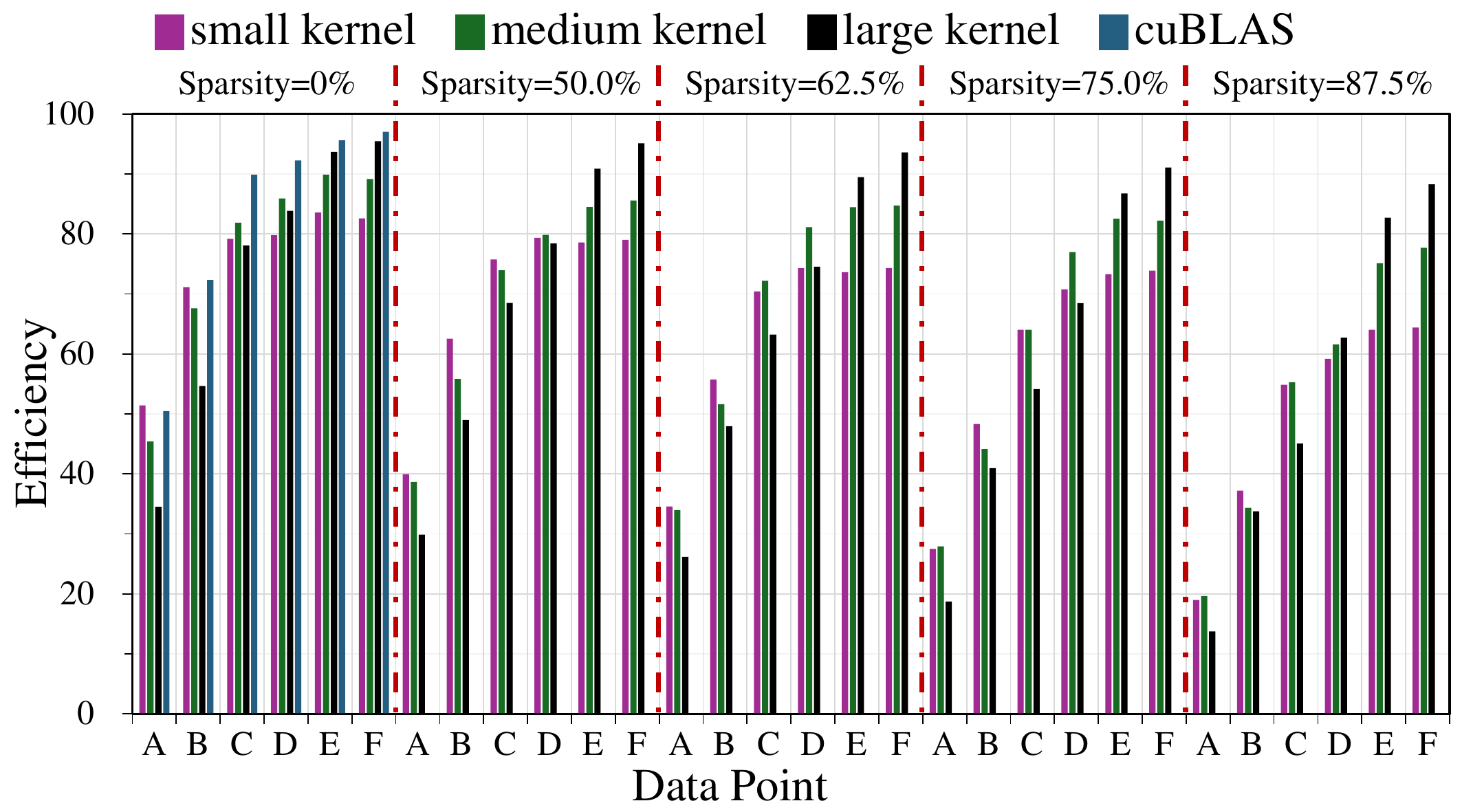}  
\caption{Performance evaluation of kernels with different blocking parameters on A100. Detailed data points can be found in Table \ref{tab:matrix_sizes}.} 
\label{fig:different_kernel}
\end{figure}

In Section \ref{subsec:Hierarchical_Blocking_Mechanism}, we propose a hierarchical blocking mechanism tailored to the characteristics of \textit{N:M} sparsity and provide recommended parameter configurations for small, medium, and large input matrices. 
As shown in Figure \ref{fig:different_kernel}, the vertical axis represents efficiency, while the horizontal axis is divided into five regions corresponding to sparsity levels of 0.0\%, 50.0\%, 62.5\%, 75.0\%, and 87.5\%. 
Each region contains six data points derived from the small, medium, and large matrices listed in Table \ref{tab:matrix_sizes}. 
The cuBLAS only appears when the sparsity level is 0.0\%, as it can only perform dense GEMM operations.

At a sparsity level of 0.0\%, our kernel nearly matches the performance of cuBLAS kernels across various input matrices of different size, even when accounting for the overhead of indirect memory access.

It is evident that, across all sparsity levels, kernels optimized for matrices with specific characteristics consistently achieve the best performance for those cases. For example, with smaller input matrices, the small kernel achieves the best performance, while the large kernel excels with larger matrices.
This highlights both the effectiveness and necessity of designing kernels with different blocking parameters tailored to various matrix characteristics.

\subsection{Performance Evaluation with Related Works}

    
\begin{figure}[htbp]
    \centering
    \captionsetup[subfigure]{skip=1pt} 
    \subfloat[Sparsity Ratio = 50.0\%]{
        \includegraphics[width=1.0\linewidth]{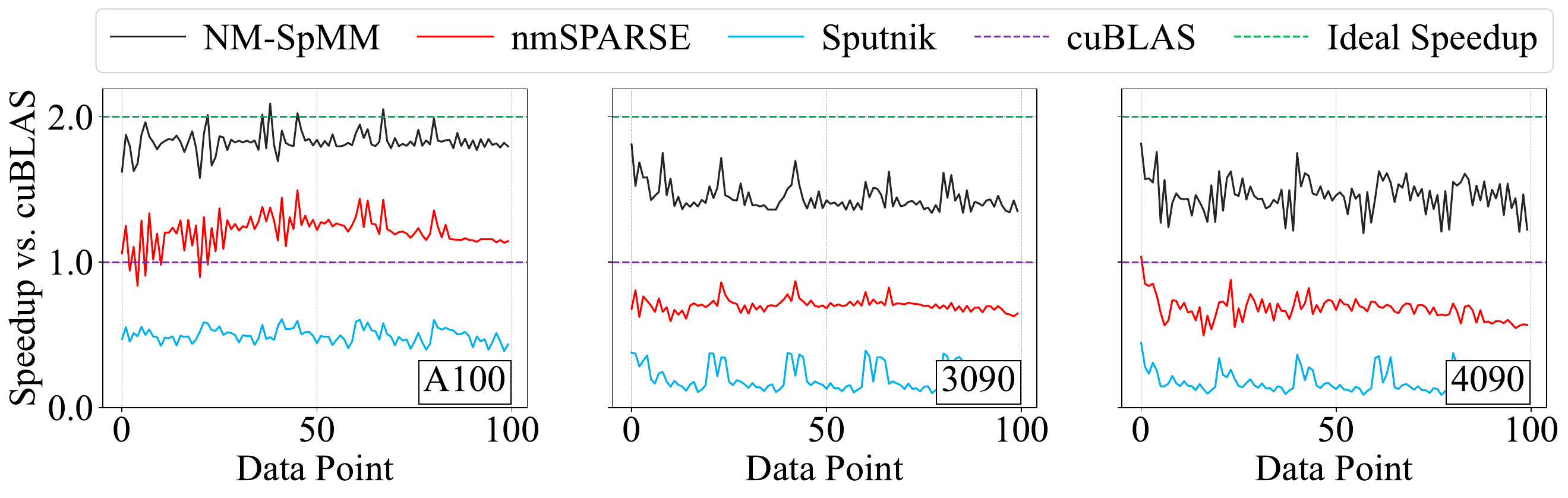}
    }
    \vspace{2pt} 
    \subfloat[Sparsity Ratio = 62.5\%]{
        \includegraphics[width=1.0\linewidth]{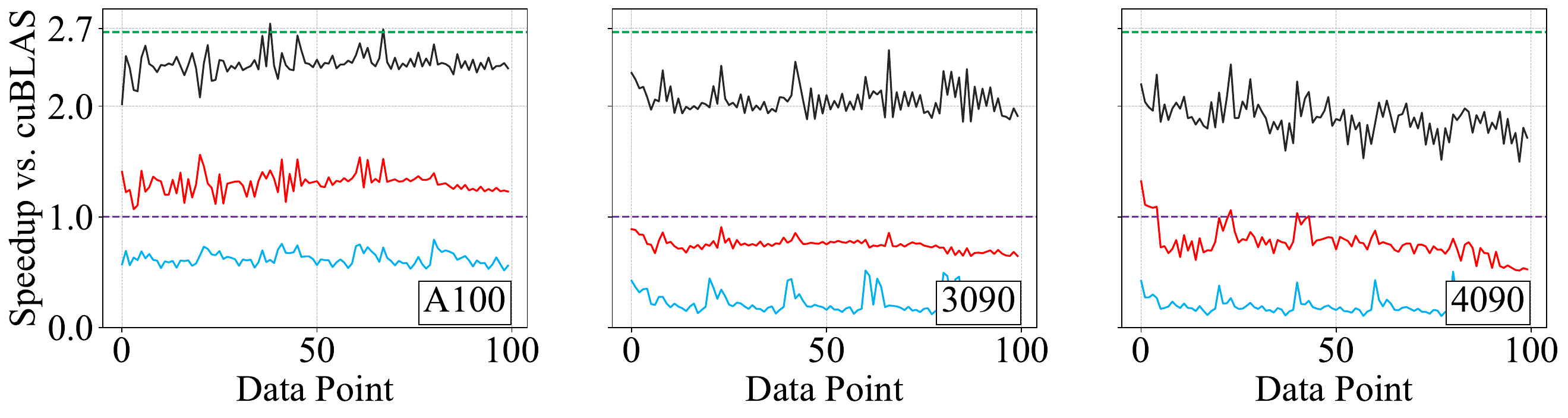}
    } 
    \vspace{2pt} 
    \subfloat[Sparsity Ratio = 75.0\%]{
        \includegraphics[width=1.0\linewidth]{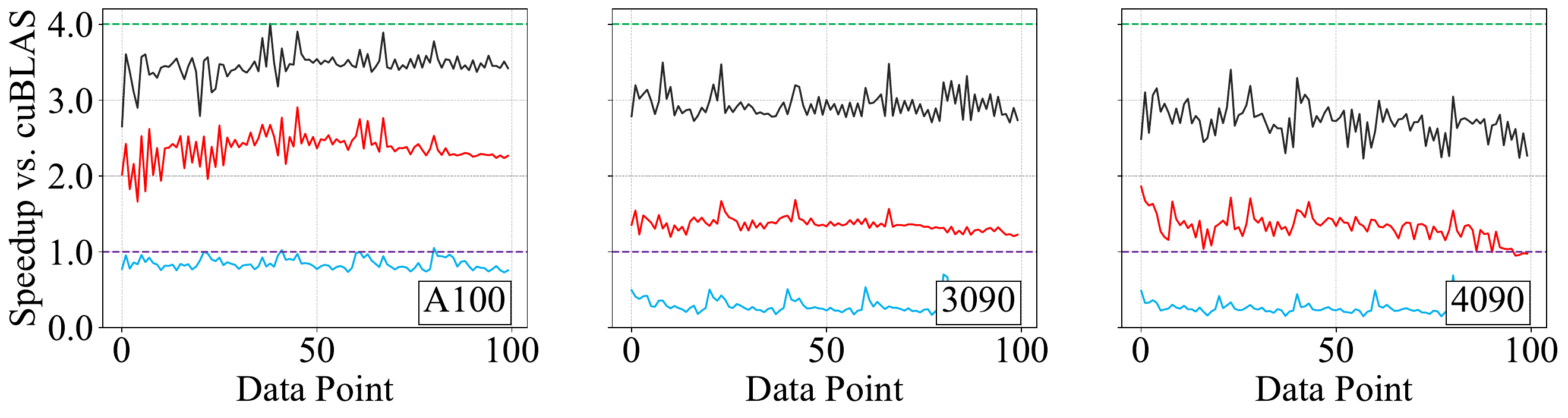}
    }
    \vspace{2pt} 
    \subfloat[Sparsity Ratio = 87.5\%]{
        \includegraphics[width=1.0\linewidth]{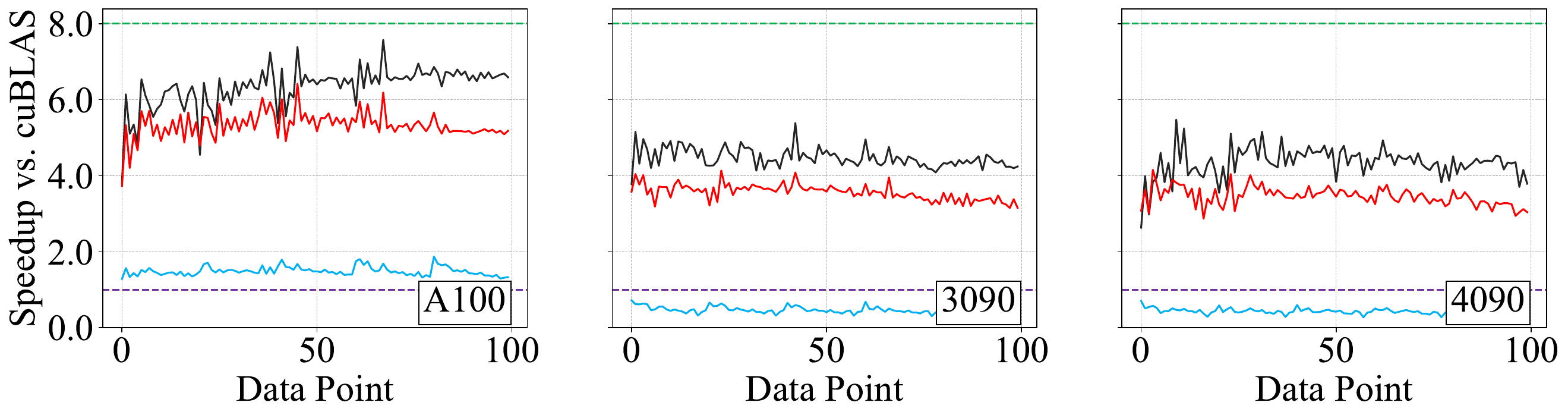}
    }
    \vspace{2pt} 
    \caption{Kernel performance On A100, 3090 and 4090.}
    \label{fig:kernel_perfomance}
\end{figure}

In this experiment, we compare NM-SpMM with three other state-of-the-art works including cuBLAS, Sputnik, and nmSPARSE on three GPU cards. The data set used in these experiments contains 100 data points described previously. Again, four different sparsity levels, with two moderate sparsity and two high sparsity, are used. The evaluation results are illustrated in Figure~\ref{fig:kernel_perfomance}.

In Figure~\ref{fig:kernel_perfomance}, the vertical axis shows the speedup relative to cuBLAS dense computation, while the horizontal axis indicates the index of 100 data points. cuBLAS serves as a baseline, marked by a constant purple dashed line at 1 across all figures. The green dashed line represents the ideal speedup achievable with sparse matrices. For instance, with 75.0\% sparsity, computation reduces to a quarter of the original, yielding an expected speedup of 4.

Similar trends are observed across various GPU cards, so we focus our analysis primarily on the data from the A100. The blue solid line represents Sputnik, which shows poorer performance due to its direct handling of unstructured sparse patterns, leading to irregular memory access and imbalanced workload overhead.

The nmSPARSE is denoted by a red solid line. It is a library designed specifically for \textit{N:M} sparsity. Compared to Sputnik using an unstructured pattern, nmSPARSE demonstrates a significant performance advantage with its \textit{N:M} sparsity pattern. Therefore, nmSPARSE achieves speedups of about 1.2x, 1.3x, 2.4x, and 5.3x over cuBLAS at sparsity levels of 50.0\%, 62.5\%, 75.0\%, and 87.5\%, respectively. However, nmSPARSE falls short in optimizing data locality and lacks sparsity-aware memory enhancements, leaving room for further improvements with our approach.

The NM-SpMM is plotted with a black solid line. Besides the \textit{N:M} sparsity pattern, NM-SpMM leverages hierarchical block algorithms to utilize data locality and applies sparsity-aware optimization.
Finally, NM-SpMM outperforms other methods. For instance, at sparsity levels of 50.0\%, 62.5\%, 75.0\%, and 87.5\%, our work achieves speedups of 1.8x, 2.4x, 3.5x, and 6.3x over cuBLAS, and speedups of 1.5x, 1.8x, 1.5x, and 1.2x over nmSPARSE, respectively.

On the 3090 and 4090, where there's a larger gap between SM computing power and memory bandwidth (see Table \ref{tab:gpu-comparison}), NM-SpMM shows smaller performance gains from \textit{N:M} sparsity but still surpasses other methods. Overall, NM-SpMM is 2.1x faster than nmSPARSE, with speedup over cuBLAS ranging from 1.4x to 6.3x.

\subsection{Roofline Analysis}
 
\begin{figure}[t]
\centering 
\includegraphics[width=1.0\linewidth]{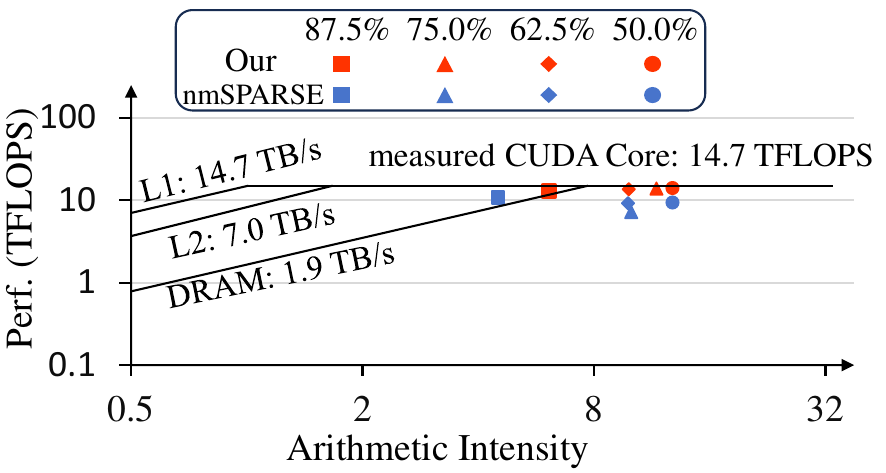}  
\caption{Roofline analysis on A100, here $m = n = k = 4096$.}
\label{fig:roofline}
\end{figure}

We conduct a roofline analysis on NM-SpMM with an input matrix with dimensions $m=n=k=4096$ on the A100 platform. 
The experimental results is illustrated in Figure~\ref{fig:roofline}. 
We select four representative sparsity levels.
In the figure, the TFLOPS on the vertical axis were collected using NVIDIA's Nsight Compute(NCU), while the arithmetic intensity on the horizontal axis was calculated using Equation \ref{eq:AI_NM_sparsity}.
The NCU locks the SM frequency to a specific value during profiling, resulting in a measured peak performance of 14.7 TFLOPS for FP32 CUDA cores.
NM-SpMM (indicated by the red markers) achieve 96\%, 93\%, 95\%, and 88\% of peak performance at sparsity levels of 50.0\%, 62.5\%, 75\%, and 87.5\%, respectively. 
In contrast, nmSPARSE, represented by the blue markers, only reach 64\%, 63\%, 49\%, and 73\%.
At sparsity levels of 75.0\% and 87.5\%, NM-SpMM's optimization to reduce memory footprint results in a higher arithmetic intensity compared to nmSPARSE.
At sparsity levels of 50.0\% and 62.5\%, the limited capacity of shared memory prevents the use of larger $k_s$ and $w_s$. Larger $k_s$ and $w_s$ could lead to higher arithmetic intensity, ultimately resulting in NM-SpMM having a higher arithmetic intensity at a sparsity level of 75.0\% compared to 62.5\%.
This experiment highlights the superiority of \textit{N:M} sparsity in hardware computation and validates the effectiveness of our optimization measures.

\section{Conclusion}
\label{sec:conclusion}

In conclusion, \textit{N:M} sparsity has proven to be a highly effective strategy to improve DNN inference performance and reduce model size. However, current GPU implementations for \textit{N:M} sparsity either lack generality or fail to deliver optimal performance. In this work, we propose NM-SpMM, a novel approach that efficiently implements vector-wise \textit{N:M} sparsity. By incorporating hierarchical blocking mechanism and sparsity-aware optimization, NM-SpMM significantly outperforms existing solutions. Our experiments show that NM-SpMM achieves a speedup of 2.1x compared to nmSPARSE and delivers speedups ranging from 1.4x to 6.3x over cuBLAS. This performance approaches the theoretical maximum speedup enabled by sparsity. This advancement underscores the potential of \textit{N:M} sparsity to enhance performance in deep learning applications.

\section{Acknowledgment}
\label{sec:conclusion}

This work was partly supported by the Shenzhen-HongKong Joint Funding Project (Category A) under Grant No. SGDX20230116092056010, Shenzhen Key Laboratory of Intelligent Bioinformatics under Grant No. ZDSYS20220422103800001.

This work was also supported by Shanghai Zelixir Biotech Company by its Joint Lab of Zelixir-SIAT, and Tencent with 3 years continuous funding support.

\bibliography{bibliography}
\bibliographystyle{IEEEtran}



    

\end{document}